\documentclass[aps,prl,twocolumn,groupedaddress,letterpaper]{revtex4}
\usepackage{graphicx}
\usepackage{amsmath}
\usepackage{esint}
\usepackage{verbatim}
\usepackage{color}
\usepackage{SIunits}
\usepackage{hyperref}
\renewcommand{\Re}{\text{Re}}
\begin{document}
\title{Optimized optomechanical crystal cavity with acoustic radiation shield}
\author{Jasper Chan}
\author{Amir H. Safavi-Naeini}
\author{Jeff Hill}
\author{Se\'{a}n Meenehan}
\author{Oskar Painter}
\email{opainter@caltech.edu}
\affiliation{Thomas J. Watson, Sr., Laboratory of Applied Physics, California Institute of Technology, Pasadena, CA 91125}
\date{\today}
\begin{abstract}
We present the design of an optomechanical crystal nanobeam cavity that combines finite-element simulation with numerical optimization, and considers the optomechanical coupling arising from both moving dielectric boundaries and the photo-elastic effect. Applying this methodology results in a nanobeam with an experimentally realized intrinsic optical $Q$-factor of $1.2\times10^6$, a mechanical frequency of $5.1$~GHz, a mechanical $Q$-factor of $6.8\times10^5$ (at $T=10$~K), and a zero-point-motion optomechanical coupling rate of $g=1.1$~MHz.
\end{abstract}
\pacs{}
\maketitle

The use of radiation pressure forces to control and measure the mechanical motion of engineered micro- and nanomechanical objects has recently drawn significant attention in fields as diverse as photonics~\cite{Li2008}, precision measurement~\cite{Regal2008}, and quantum information science~\cite{Stannigel2010}.   A milestone of sorts in cavity circuit- and optomechanics is the recent~\cite{Teufel2011b,Chan2011} cooling of a mechanical resonator to a phonon occupancy $\langle n \rangle \lesssim 1$ using cavity-assisted radiation pressure backaction~\cite{Braginsky1977,Marquardt2007}.  Backaction cooling involves the use of an electromagnetic cavity with resonance frequency $\omega_o$ sensitive to the mechanical displacement, $x$, of the mechanical resonator.  The canonical system is a Fabry-Perot cavity of length $L$ with one end mirror fixed and with the other end mirror of mass $m_\text{eff}$ mounted on a spring with resonance frequency $\omega_m$.  The coupling between the electromagnetic field and mechanics is quantified by the frequency shift imparted by the zero-point motion of the mechanical resonator, given by $g = g_\text{OM}\sqrt{\hbar/2m_\text{eff}\omega_m}$, where $g_\text{OM} = \partial\omega_o/\partial x = \omega_o/L$. 


One of many technologies recently developed to make use of radiation pressure effects are optomechanical crystals (OMCs)~\cite{Kang2009,Eichenfield2009b}.  Optomechanical crystals, in their most general form, are quasi-periodic nanostructures in which the propagation and coupling of optical and acoustic waves can be engineered.  In this work we present the comprehensive design, fabrication, and characterization of a quasi-1D OMC cavity formed from the silicon device layer of a silicon-on-insulator (SOI) microchip.  Our design incorporates both moving-boundary and photo-elastic (electrostriction) radiation pressure contributions, and simultaneously optimizes for optical and acoustic parameters.  



\begin{figure}[btp]
\begin{center}
\includegraphics[width=\columnwidth]{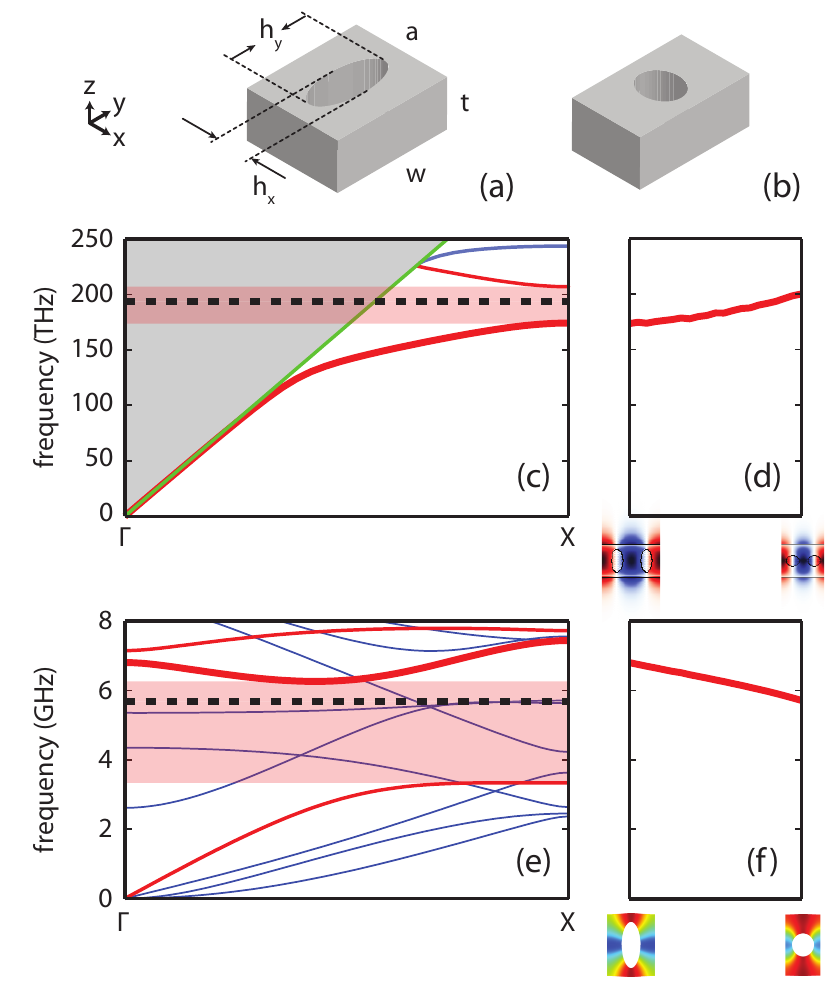}
\caption{ The (a) nominal unit cell with $(a,t,w,h_x,h_y) = (436,220,529,165,366)$~nm and (b) defect unit cell with $(a,t,w,h_x,h_y) = (327,220,529,199,170)$~nm of the OMC nanobeam cavity. The (c) optical and (e) mechanical band structure for propagation along the $x$-axis in the nominal unit cell, with quasi-bandgaps (red regions) and cavity mode frequencies (black dashed) indicated. In (c), the light line (green curve) divides the diagram into two regions: the gray shaded region above representing a continuum of radiation and leaky modes, and the white region below containing guided modes with $y$-symmetric (red bands) and $y$-antisymmetric (blue bands) vector symmetries. In (e), modes that are $y$- and $z$-symmetric (red bands), and modes of other vector symmetries (blue bands) are indicated.  The bands from which the localized cavity modes are formed are shown as thicker curves. Tuning of the (d) $X$-point optical and (f) $\Gamma$-point mechanical modes of interest as the unit cell is smoothly transformed from the nominal to the defect unit cell.} \label{fig:unit_cell}
\end{center} 
\end{figure}

The nominal unit cell of a nanobeam OMC, geometrically a silicon block with an oval hole in it, is shown schematically in Fig.~\ref{fig:unit_cell}a.  The corresponding optical and mechanical bandstructure diagrams are shown in Figs.~\ref{fig:unit_cell}c and d, respectively. As indicated by the gray shaded region in the photonic bandstructure, the continuum of unguided optical modes above the light line precludes the existence of a complete photonic band gap (only a quasi-bandgap exists for the guided modes of the beam).  The physical dimension of the unit cell block (see caption of Fig.~\ref{fig:unit_cell}a) are chosen to yield a photonic quasi-bandgap surrounding a wavelength $\lambda\sim1550$~nm.  The corresponding mechanical bandstructure has a series of acoustic bands in the GHz frequency range (the ratio of the optical frequencies to that of the mechanical frequencies is roughly the ratio of the speed of light to sound in silicon).  By classifying the acoustic bands by their vector symmetry, we can again define a quasi-bandgap for the mechanical system in terms of modes of common symmetry (modes of different symmetry may couple weakly due to symmetry breaking introduced by the fabrication process, an issue we discuss and mitigate later in the design process). 

Using previously developed design intuition~\cite{Eichenfield2009b}, we choose to focus on the optical ``dielectric'' band at the $X$-point and the mechanical ``breathing mode'' band at the $\Gamma$-point (these bands emphasized by thicker lines in Figs.~\ref{fig:unit_cell}c and e).  Fig.\ref{fig:unit_cell}d and f shows the optical dielectric band $X$-point and the breathing mode $\Gamma$-point frequency as the unit cell is transformed smoothly from that in Fig.~\ref{fig:unit_cell}a to that in Fig.~\ref{fig:unit_cell}b.  We see that such a transition---simultaneously reducing the lattice constant $a$ and decreasing the hole aspect ratio, $r_h \equiv h_y/h_x$---causes an increase in the $X$-point optical frequency and a reduction in the $\Gamma$-point mechanical frequency, pushing both optical and mechanical modes into the quasi-bandgap of their respective bands (shown as shaded regions in Figs.~\ref{fig:unit_cell}c and e).



\begin{figure}[tp]
\begin{center}
\includegraphics[width=\columnwidth]{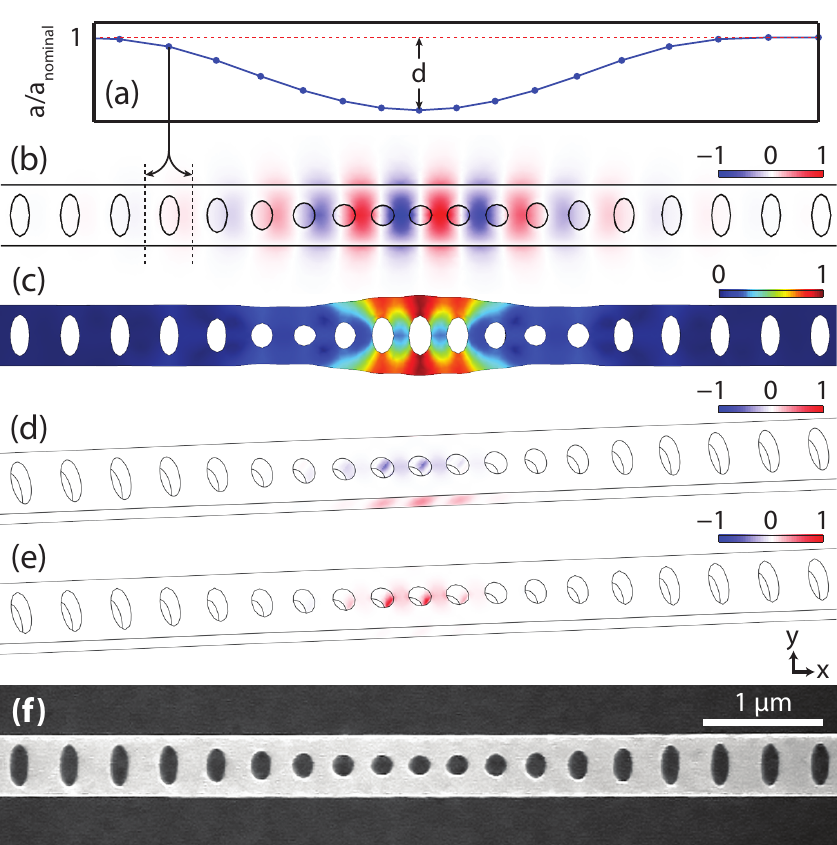}
\caption{(a) Plot of the unit cell lattice constant, $a$, along the length of the nanobeam. (b) The normalized optical $E_y$ field and (c) the normalized mechanical displacement field $|\mathbf{Q}|$ of the localized optical and mechanical modes, respectively. (d) The normalized surface density of the integrand in eq. (\ref{eqn:mb}), showing the contributions to $g_\text{OM,MB}$. (e) The normalized volumetric density of the integrand in eq. (\ref{eqn:pe}), showing the contributions to $g_\text{OM,PE}$. (f) Scanning electron microscope (SEM) image of the experimentally realized cavity. } \label{fig:cavity}
\end{center}
\end{figure}

An optomechanical cavity can be formed by transitioning from the nominal unit cell in Fig.~\ref{fig:unit_cell}a, forming the ``mirror'' region, to the defect unit cell of Fig.~\ref{fig:unit_cell}b.  This cavity can be parameterized by the maximum change in $a$ (defined as $d \equiv \max\{1-a/a_\text{nominal}\}$), the number of holes in the defect region, the maximum curvature of $a/a_\text{nominal}$ (plotted in Fig.\ref{fig:cavity}a) and the $r_h$ of the center hole. Including the unit cell geometric parameters $a$, $w$, $h_x$, and $h_y$ ($t$ is fixed to 220~nm), an entire nanobeam design is specified by these 8 values. Finite-element-method (FEM) simulations of a complete structure are used to determine the fundamental cavity mode frequencies ($\omega_o$ and $\omega_m$), motional mass $m_\text{eff}$~\cite{Safavi-Naeini2010a}, and radiation-limited optical $Q$-factor, $Q_o$. To quantify the coupling rate, we consider both the frequency shift due to the moving dielectric boundary~\cite{Johnson2002} and the photo-elastic effect~\cite{Biegelsen1974}, so that $g_\text{OM} = g_\text{OM,MB} + g_\text{OM,PE}$. The moving boundary contribution is given by~\cite{Safavi-Naeini2010a},
\begin{equation}\label{eqn:mb}
g_\text{OM,MB} = -\frac{\omega_o}{2}\frac{\oint{(\mathbf{Q}\cdot\hat{\mathbf{n}})(\Delta\varepsilon\mathbf{E}_{||}^2 - \Delta\varepsilon^{-1}\mathbf{D}_{\perp}^2)}\,dS}{\int{\mathbf{E}\cdot\mathbf{D}\,dV}},
\end{equation}
where $\mathbf{Q}$ is the normalized displacement field ($\max\{|\mathbf{Q}|\} = 1$), $\hat{\mathbf{n}}$ is the outward facing surface normal, $\mathbf{E}$ is the electric field, $\mathbf{D}$ is the displacement field, the subscripts $||$ and $\perp$ indicate the field components parallel and perpendicular to the surface, respectively, $\varepsilon$ is the material permittivity, $\Delta\varepsilon \equiv \varepsilon_\text{silicon}-\varepsilon_\text{air}$, and $\Delta\varepsilon^{-1} \equiv \varepsilon_\text{silicon}^{-1}-\varepsilon_\text{air}^{-1}$~\cite{Johnson2002}. A similar result can be derived for the photo-elastic contribution from first-order perturbation theory,
\begin{equation}
g_\text{OM,PE} = -\frac{\omega_o}{2}\frac{\langle E |\frac{\partial\varepsilon}{\partial\alpha}|E\rangle}{\int{\mathbf{E}\cdot\mathbf{D}\,dV}},
\end{equation}
where $\alpha$ is a generalized coordinate parameterizing the amplitude of $\mathbf{Q}$. In an isotropic medium with refractive index $n$, we have $\frac{\partial\varepsilon_{ij}}{\partial\alpha} = -\varepsilon_0n^4p_{ijkl}S_{kl}$ where $\mathbf{p}$ is the rank-four photo-elastic tensor and $\mathbf{S}$ is the strain tensor~\cite{Yariv1983}. For silicon, a cubic crystal with point symmetry group m3m, with the $x$-axis and $y$-axis respectively aligned to the $[100]$ and $[010]$ crystal direction, we can use the contracted index notation with some simplification to write
\begin{align}\label{eqn:pe}
\langle E|\tfrac{\partial\varepsilon}{\partial\alpha} |E\rangle =& -\varepsilon_0n^4\int\Big[2\Re\{E_x^*E_y\}p_{44}S_4 \nonumber\\
&+ 2\Re\{E_x^*E_z\}p_{44}S_5+\Re\{E_y^*E_z\}p_{44}S_6\nonumber\\
&+ |E_x|^2(p_{11}S_1 + p_{12}(S_2+S_3)) \nonumber\\
&+ |E_y|^2(p_{11}S_2 + p_{12}(S_1+S_3))\nonumber\\
&+ |E_z|^2(p_{11}S_3 + p_{12}(S_1+S_2)) \Big] \, dV 
\end{align}
where $(p_{11},p_{12},p_{44})=(-0.094,0.017,-0.051)$ \cite{Yariv1983,Biegelsen1974}. 

In order to optimize the nanobeam OMC design we have chosen to assign a fitness value $F \equiv -g\cdot\min\{Q_o,Q_\text{cutoff}\}/Q_\text{cutoff}$ to the simulations.  The additional $Q_\text{cutoff}$ term (set to $3\times10^6$) is used to avoid unrealizably high simulated radiation-limited $Q_o$ values from unfairly weighting the fitness.  With this choice the nanobeam design has been reduced to an 8 parameter optimization problem amenable to a variety of numerical minimization techniques. For a computationally expensive fitness function with a large parameter space, a good choice of optimization algorithm is the Nelder-Mead method~\cite{Lagarias1998}; as a downhill simplex method (compared to a gradient descent method) the search technique is resistant to simulation noise and discontinuities. To mitigate the problem of converging on local minima, the optimization procedure is applied to many randomly generated initial conditions. The resulting optimized nanobeam OMC design is shown in Fig.~\ref{fig:cavity}, with a simulated $\omega_o/2\pi$ of $194$~THz, radiation-limited $Q_o$ of $2.2\times10^7$, $\omega_m/2\pi$ of $5.7$~GHz, and $m_\text{eff}$ of $127$~fg. The total optomechanical coupling rate, $g/2\pi$, is 770~kHz, composed of a $-90$~kHz contribution from the moving boundary and a $880$~kHz contribution from the photo-elastic effect.

\begin{figure}[btp]
\begin{center}
\includegraphics[width=\columnwidth]{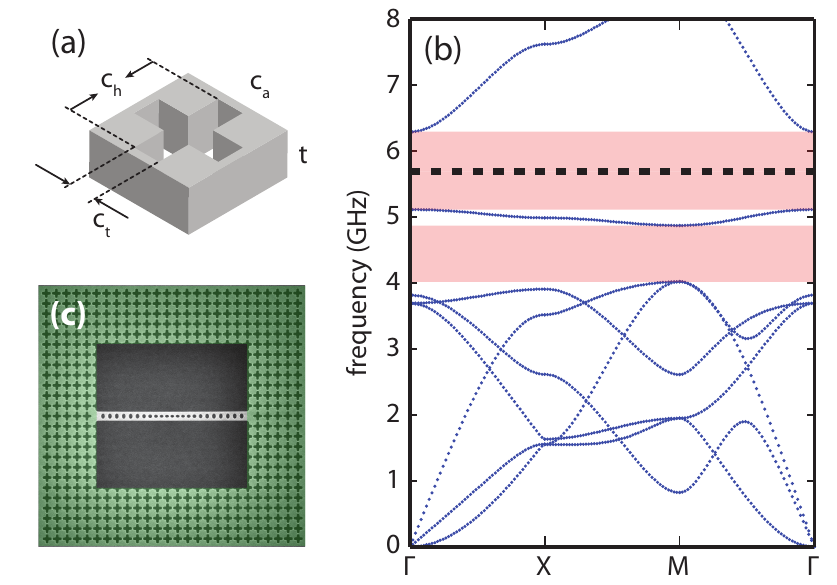}
\caption{(a) The unit cell of the phononic shield with parameters $(c_a,c_h,c_t,t) = (534,454,134,220)$~nm. (b) Full in-plane mechanical band diagram with complete phononic bandgap (red region) and frequency of the acoustic mode of the cavity indicated (black dashed line). (c) SEM image showing the phononic shield (green) around the nanobeam.} \label{fig:phononic_shield}
\end{center}
\end{figure}

Unaddressed so far in our design is mechanical losses, a source of which is coupling (made unavoidable by inevitable symmetry breaking during the fabrication process) to unconfined modes of alternate symmetries present in the quasi-bandgap of the patterned nanobeam (Fig.~\ref{fig:unit_cell}e).  These acoustic radiation losses can be significantly suppressed by surrounding the entire nanobeam inside a second acoustic shield consisting of a patterning with a complete phononic band gap. The 2D cross phononic crystal design~\cite{Safavi-Naeini2010a} has previously demonstrated this to great effect~\cite{Alegre2011}, and the wide tuneability of the acoustic gap essentially allows the mechanical loss engineering to be decoupled from the design of the co-localized cavity modes of the nanobeam. By adjusting $c_a$,  $c_h$ and $c_t$ in the 2D unit cell (Fig.~\ref{fig:phononic_shield}a), it can be seen from Fig.\ref{fig:phononic_shield}b that the complete bandgap can be tailored to be centered on $\omega_m$.

The optimized nanobeam design was fabricated from the [001] silicon device layer of a silicon-on-insulator wafer from SOITEC (resistivity $4\mathord{-}20$~$\Omega\cdot$cm, device layer thickness $220$~nm, buried-oxide layer thickness $2\mathord{-}3$~\micro{m}). The cavity geometry was defined by electron beam lithography followed by inductively-coupled-plasma reactive-ion etching to transfer the pattern through the $220~\text{nm}$ silicon device layer. The cavities were then undercut using a 1:1 HF:H$_2$O solution to remove the buried oxide layer, and cleaned using a piranha/HF cycle. The silicon surface was hydrogen-terminated with a weak 1:20 HF:H$_2$O solution for chemical passivation~\cite{Borselli2006}.

\begin{figure}[btp]
\begin{center}
\includegraphics[width=\columnwidth]{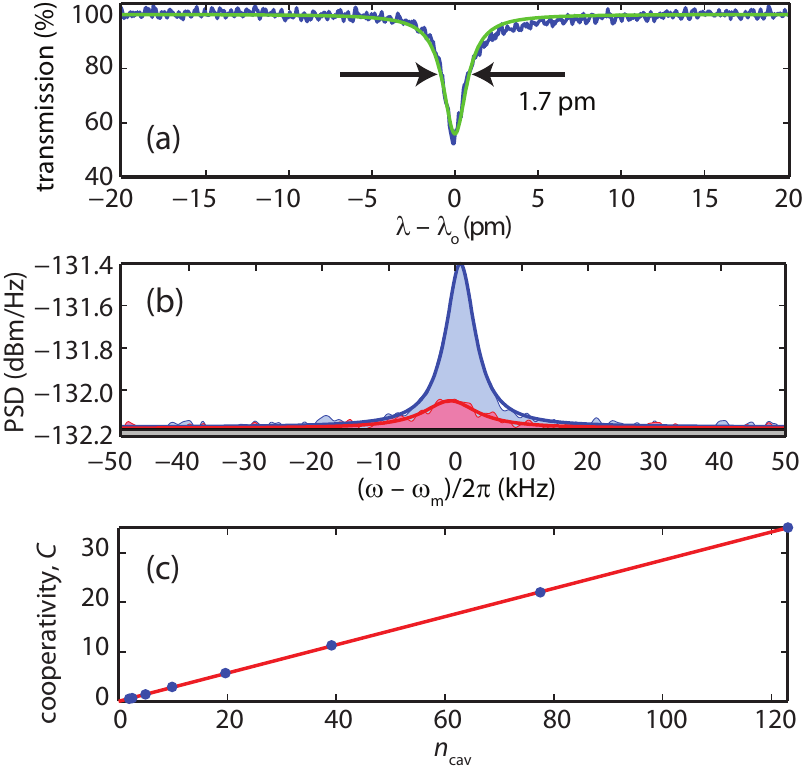}
\caption{(a) Normalized optical transmission spectrum, centered at $1544.8$~nm, showing the fundamental optical cavity mode of the nanobeam, with a measured intrinsic $Q_{o,i}=1.22\times10^6$. (b) Optically transduced thermal noise power spectral density centered at the mechanical frequency, $\omega_m/2\pi = 5.1$~GHz, of the breathing mode, taken at $T_s=10$~K with the input laser red-detuned ($\Delta=\omega_m$; red curve) and blue-detuned ($\Delta=-\omega_m$; blue curve) from the cavity.  (c) Measured cooperativity, $C$, as function of intracavity photon number, $n_\text{cav}$, for red detuning $\Delta=\omega_m$.} \label{fig:exp}
\end{center}
\end{figure}

Characterization of the OMC nanobeam devices was performed in a continuous-flow helium cryostat, under vacuum and at a sample mount temperature of $T_s \approx 6$~K.  A tapered optical fiber~\cite{Michael2007}, positioned in the optical near-field ($\mathord\sim100$~nm) using a set of low-temperature-compatible piezoelectric stages, is used to evanescently couple laser light into and out of the devices.  A tunable external cavity diode laser (New Focus, model 6728) is used to scan across the wavelength band from $\lambda=1520\mathord{-}1570$~nm.  The resulting normalized optical transmission scan of an OMC nanobeam cavity with resonance wavelength at $\lambda_o=1544.8$~nm ($\omega_o/2\pi=194$~THz) is shown in Fig.~\ref{fig:exp}a.  A Lorentzian fit to the measured optical linewidth ($\delta_{\lambda}=1.7$~pm) yields a fiber-taper-loaded optical $Q$-factor of $Q_o=9.06\times10^5$, which for the measured on-resonance transmission of $55\%$ corresponds to an intrinsic $Q$-factor of $Q_{o,i}=1.22\times10^6$.  The corresponding total cavity (energy) decay rate, (bi-directional) fiber taper waveguide coupling rate, and intrinsic decay rate are $\kappa/2\pi=214$~MHz, $\kappa_e/2\pi=55$~MHz, and $\kappa_i/2\pi=159$~MHz, respectively.     

Spectroscopy of the mechanical mode is performed by tuning the frequency of the input laser ($\omega_l$) to either a mechanical frequency red- or blue-detuned from the optical cavity resonance ($\Delta\equiv(\omega_o-\omega_l)=\pm\omega_m$).  The optomechanical backaction under such conditions results in an optically induced damping of $\gamma_{\text{OM}}=\pm4G^2/\kappa$ for $\Delta=\pm\omega_m$~\cite{Safavi-Naeini2011}, where $G=\sqrt{n_\text{cav}}g$ is the parametrically-enhanced optomechanical coupling.  The intracavity photon number can be related to the laser input power ($P_i$) through the relation, $n_\text{cav}=P_i\left(\kappa_e/2\hbar\omega_l((\kappa/2)^2+\Delta^2)\right)$.  The thermal Brownian motion of the mechanical mode is imprinted on the transmitted laser light as a sideband of the input laser resonant with the optical cavity resonance.  As shown in Fig.~\ref{fig:exp}b, the resulting intensity modulation of the photodetected signal gives rise to a Lorentzian response in the photocurrent electronic power spectrum around the $5.1$~GHz resonance frequency of the breathing mechanical mode.  The intrinsic mechanical damping of the breathing mode is found by averaging the measured mechanical linewidths under red and blue detuning, $\gamma_{i}=(\gamma_+ + \gamma_-)/2 = 7.5$~kHz, where $\gamma_\pm=\gamma_i\pm\gamma_{\text{OM}}$.  The corresponding intrinsic mechanical $Q$-factor is $Q_{m,i}=6.8\times10^5$.  Experiments with different surface preparations (with and without HF-dip prior to testing) indicate that $Q_{m,i}$ is extremely sensitive to the surface quality, and is likely not limited by bulk material damping at these temperatures~\cite{Alegre2011}.  A plot of the cooperativity, $C=\gamma_{\text{OM}}/\gamma_i=\gamma_+/\gamma_i -1$, is shown in Fig.~\ref{fig:exp}b versus $n_c$, yielding the zero-point optomechanical coupling rate of $g=1.1$~MHz for the breathing mode, very close to the simulated value (discrepancy here is attributed to the uncertainty in the silicon photo-elastic coefficients, which were extrapolated from data at wavelengths of 3.4~\micro{m} and 1.15~\micro{m}~\cite{Biegelsen1975,Biegelsen1974}).

These measured device parameters place the system in the deeply resolved sideband regime ($\omega_m/\kappa \gtrsim 30$), suitable for efficient laser cooling of the mechanical resonator into its quantum ground state.  Calibration of the thermal Brownian motion~\cite{Chan2011} at low incident laser power ($n_\text{cav}\lesssim 1$) indicates that the local bath temperature of the $5.1$~GHz breathing mode is $T_b\approx10$~K, corresponding to a thermal bath occupancy of only $n_b=43$, and a thermal decoherence time of $\tau_{\text{th}} = \hbar Q_m/k_{\text{B}}T_b \approx 0.5$~\micro{s} ($\mathord{>}10^4$ cycles of the mechanical resonator).  With a measured (simulated) granularity parameter of $g/\kappa_i\approx 0.0069$ ($0.12$), such devices with further improvement in the optical $Q$-factor, represent a promising platform for realizing quantum nonlinear optics and mechanics~\cite{Nunnenkamp2011}.          

This work was supported by the DARPA/MTO ORCHID program through a grant from AFOSR, and the Kavli Nanoscience Institute at Caltech. JC and ASN gratefully acknowledges support from NSERC.


\end{document}